# Antarctica: A Southern Hemisphere Windpower Station?

**Alexander A. Bolonkin**[•]
C & R, 1310 Avenue R, Suite 6-F
Brooklyn, New York 11229, USA
aBolonkin@juno.com,

**Richard B. Cathcart**
Geographos
1300 West Olive Avenue, Suite M
Burbank, California 91506. USA

**ABSTRACT**

The International Polar Year commences in 2007. We offer a macroproject plan to generate a large amount of electricity on the continent of Antarctica by using sail-like wind dams incorporating air turbines. Electricity can be used to make exploration and exploitation efforts on Antarctica easier. We offer the technical specifications for the Fabric Aerial Dam and indicate some of the geographical facts underpinning our macro-engineering proposal.

------------------
[•] Corresponding Author

**INTRODUCTION**

Including all air flows within a 1 km layer above the planet's solid surface, a technically possible wind energy resource base would likely be about 6 TW; probably less than 10% of that flux initially energized by the Sun could actually now be converted directly to power-line distributed electricity without causing major detrimental changes to the Earth-atmosphere (Smil, 2003). Scientific computer modeling has shown that the extraction of kinetic energy from wind would alter Earth's "…turbulent transport in the atmospheric boundary layer" (Keith et al., 2004).

Nevertheless, wind is a clean source of energy that has been utilized by humans for centuries to grind grains, pump water, propel sailing craft, and to perform other work. Artists such as Tal Streeter, Howard Woody and Tsutomu Hiroi have flown kite-like sculptures and Jose M. Yturralde flew geometrical structures; Otto Piene's "Olympic Rainbow" consisted of five 600 m-long helium-filled polyethylene tubes displayed at the 20[th] Olympiad in Munich, Germany, during 1972. Piene coined the descriptive term "Sky Art" in 1969. "Wind farm" is the popular term used for an aggregation of wind turbines clustered at a site with persistent favorable air fluxes (Hayes, 2005). Unfortunately, existing wind energy systems have deficiencies that limit their commercial applications: (1) wind energy is unevenly distributed geographically and has relatively low energy density. A single huge wind turbine cannot be placed on the ground. Instead, numerous small wind turbines must be used; (2) wind power is a function of the cube of wind velocity. At ground level, wind speed is low and rarely steady; (3) wind power system productivity is entirely dependent on prevailing weather, making it nearly impossible for productivity to be scheduled; (4) wind turbines of conventional design produce noise and are aesthetically unattractive.

## 2. DESCRIPTION

It was not until *circa* 1840 that Antarctica was established to be an isolated continent. Its coastline is about 18,000 km. Dry katabic (gravity-driven) winds blow coastward from the high interior icecap. The windiest place in Earth close to sea level is Cape Denison $(67.02^0$ S, $142.58^0$ E) in Commonwealth Bay, Antarctica, where winds exceeding 50 m/s have been recorded regularly (Parish, 1981). Of all the continents, only on Antarctica does a single meteorological element (wind) have such an overwhelming influence on the climate. The dry katabic winds blow with great constancy in direction, often moving at 20-40 m/s over the smoothest icecap surface for hundreds of kilometers. As the katabic winds leave the South Pole and approach to within about 100 km of the coastline, they tend to decrease in speed owing to drag over a rougher ice surface. Thereafter, Antarctica's winds generally blow off the continent's coastal escarpment toward the Antarctic Circle (Fig.1).

Worldwide, there are many macroproject R&D programs for the development of wind energy systems but most of them are ground or tower based; Australian macroengineers have proposed Earth-stratosphere deployed kite-like electricity generators tethered to the ground by strongly anchored cables. We propose an innovative wind energy harnessing system, the Fabric Aerial Dam (FAD), which can operate successfully up to the lower boundary of the Earth's stratosphere; our defining challenge is to manage the transfer of the energy obtained to the consuming ground-based infrastructure. We propose that the first installation be made on the continent of Antarctica because electrical power is needed and because the wind energy harvesting system we propose can cause a "windbreak" or "shelterbelt" effect downwind that would be inappropriate for a densely populated region.

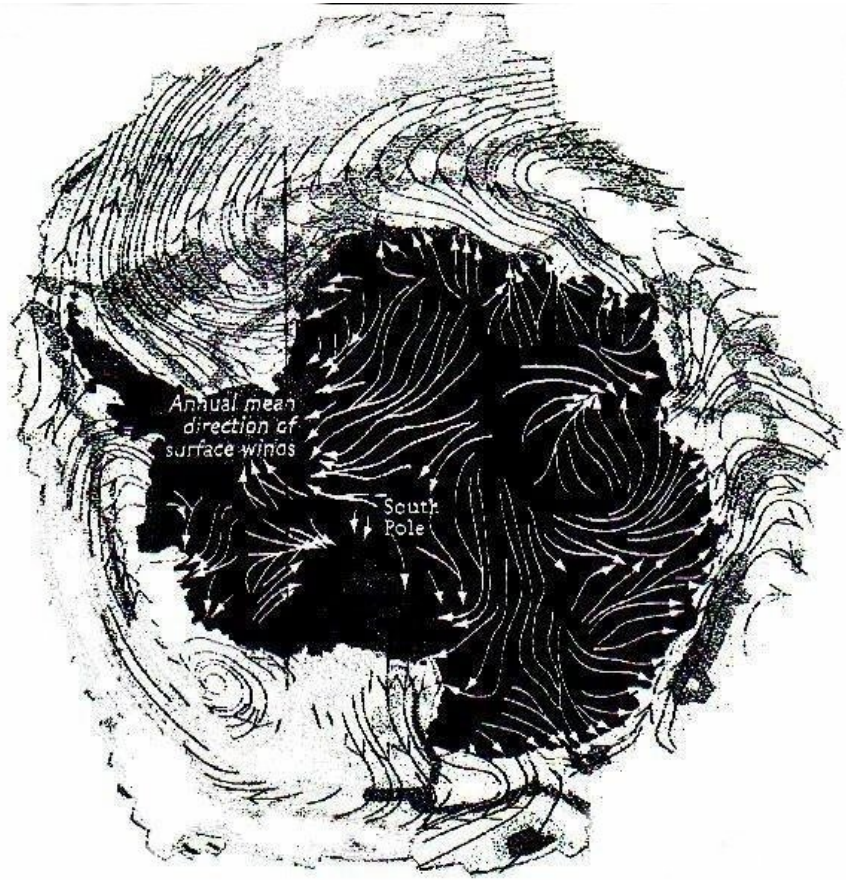

**Fig. 1.**   Map of annual mean direction of surface wind over ice.

In 1997, the first truly accurate map of Antarctica was produced by the Canadian Space Agency using its Radarsat-1 Earth-orbiting satellite (Murray, 2005).  Antarctica's winter starts in April and ends during September; at the South Pole (elevation 2835 m), the sun rises on September 21 and sets on March 21.  Antarctica has no indigenous human population; about 1,000 persons over-winter and 20,000 persons may work during summer on the icy continent.  The lowest temperature ever measured was recorded in 1983 at Russia's Vostok Station (elevation of 3744 m) in East Antarctica—minus $89^0$ C. Started in 2002, and virtually completed by 2006, Antarctic-1 is the continent's only "ice highway", connecting McMurdo on the coast to the Amundsen-Scott South Pole Base (George, 2004); Antarctic-1 was anticipated by an artist, the sculptor Rachel Weiss, more than twenty years ago (Weiss, 1984)!  FAD, the Fabric Aerial Dam can be used as a beautifying decoration, as a projection screen for nighttime motion pictures (outdoor cinema), as a nighttime solar reflector and as a daytime partly focused heater for targeted small regions on the continent of Antarctica.

The Fabric Aerial Dam (FAD) is diagramed in Fig. 2.

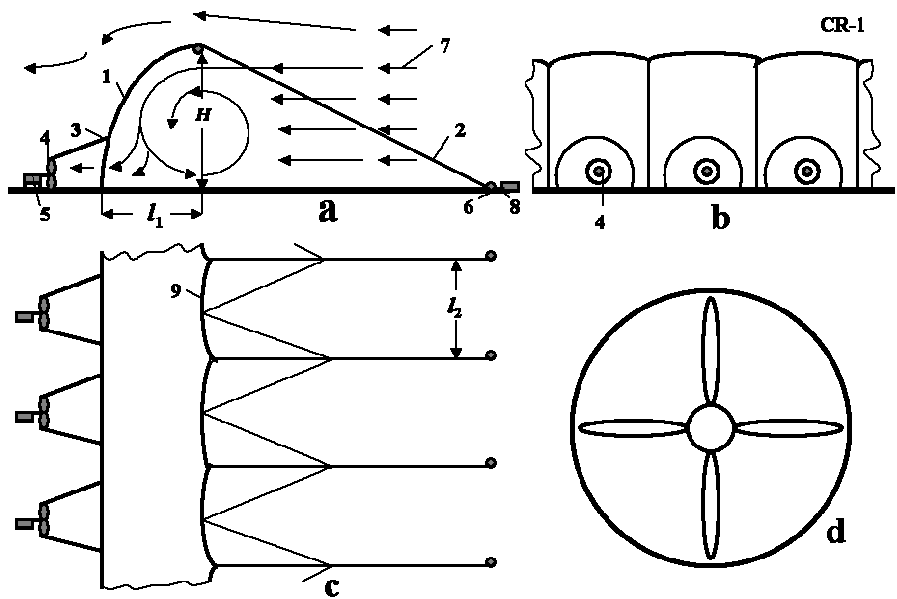

**Fig. 2.** Fabric Aerial Dam and Wind Turbine Station. (a) side-view, (b) front-view, (c) wind engine.  Notation: 1-flexible film aerial dam, 2-tethering cables, 3-air channel, 4-air turbine (propeller), 5-electricity generator, 6-support cable spool, 7-wind, 8-spool motor, 9-film cable.  $H$ - deployed elevation of FAD.

The FAD embodies a thin, possibly transparent, film 1, support cables 2, conic tunnels (3) funneling naturally flowing air to the turbines 4, electric generator 5, spool 6 for support cable, spool motor 8, film cable 9.  The FAD will be installed perpendicular to the average main direction of the site's wind at an altitude where the wind is more than 1 m/s most days of the year. The FAD shields the downwind territory from extreme wind impacts and it can help to cause orographic precipitation. If storm a wind blows in opposed direction, the spools (6) can reel in the cable (2) and settle the flexible film aerial dam material safely on the ground for temporary storage. The same technique may be used when repairs to the FAD are found to be necessary.

If the wind is less than 1m/s, the FAD will tend to fall to ground but with a stronger wind the FAD will billow, taking off from the ground; if the wind is excessively strong, however, it may irreparably damage the FAD. We can calculate the minimum and maximum admissible (safe) wind billowing a FAD. Our purpose in doing so is to estimate the time (% or actual days in a year) when the FAD can operate properly. We assume the average wind speed at an altitude of $H_0 = 10$ m and 50 m is approximately $V_0$ = 8 and 12.7 m/s respectively.

The change of wind via altitude approximately is described by equation

$$\frac{V}{V_0} = \left(\frac{H}{H_0}\right)^{\alpha} , \qquad (1)$$

where $V_0$ is the wind speed at the original height, $V$ the speed at the new height, $H_0$ the original height, $H$ the new height, and $\alpha$ the surface roughness exponent.

We assume the surface roughness exponent, $\alpha$, over Antarctica's ice is 0.10.

The result of our computations is shown in Fig. 3 below.

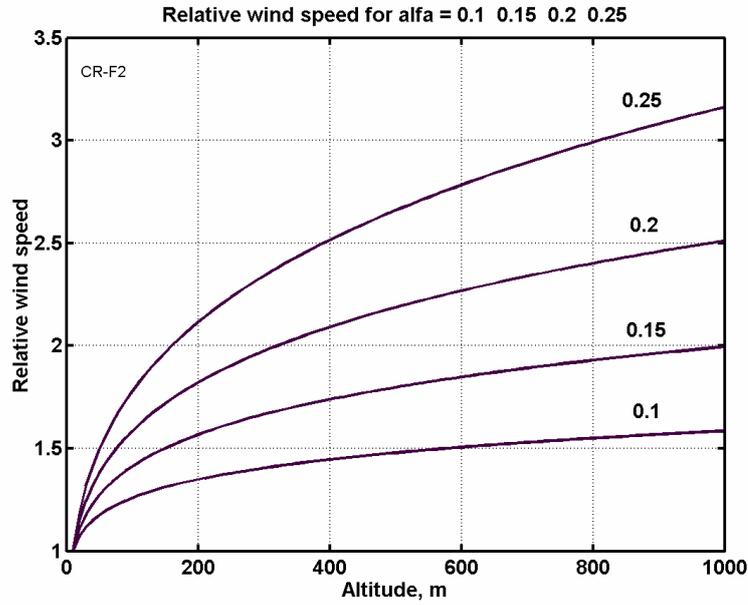

**Fig.3.** Relative wind speed via altitude
and Earth Surface. For ice $\alpha = 0.1$.

Annual speed distributions vary widely from one site to another, reflecting climatic and geographic conditions. Meteorologists have found that the Weibull probability function best approximates the distribution of wind speeds over time at sites around the world where actual distributions of wind speeds are unavailable. The Rayleigh distribution is a special case of the Weibull function, requiring only the average speed to define the shape of the distribution.

Equation of Rayleigh distribution is

$$f_x(x) = \frac{x}{\alpha^2}\exp\left[-\frac{1}{2}\left(\frac{x}{\alpha}\right)^2\right], \quad x \geq 0, \quad E(X) = \sqrt{\frac{\pi}{2}}\alpha, \quad Var(X) = \left(2 - \frac{\pi}{2}\right)\alpha^2, \quad (2)$$

where $\alpha$ is new parameter.

These data gives possibility to easy calculate the amount (percent) days (time) when air dam can operate in year (fig.4). It is very important value for the estimation efficiency of offered devices.

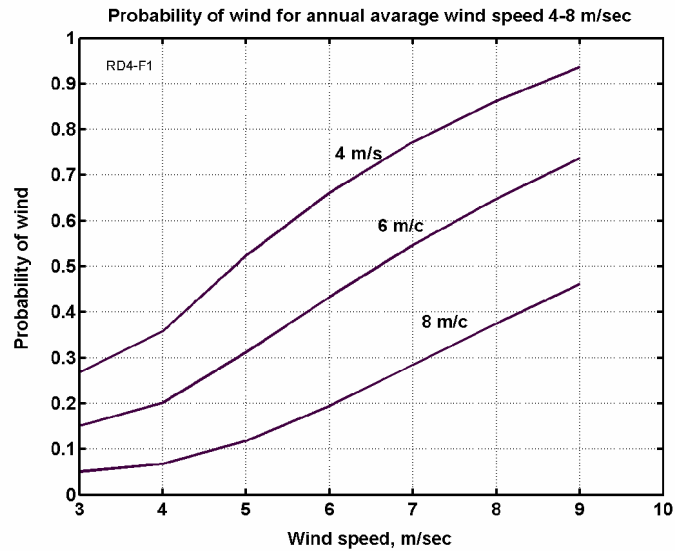

**Fig. 4.** Probability of wind via wind speed and average wind speed in given place.

  Let us compute two examples:

1)  Assume, the air dam has minimum admissible wind speed 3 m/s, the average annual speed in given region is 6 m/s. From fig.4, Eq. (2) , we can get the probability wind less 3 m/sec is 15%. The same way we can compute the probability of a storm wind speed.

2)  Assume, the average annual speed in given region is 6 m/s, the maximum admissible wind speed is 7 m/s. The probability that a wind speed will be less then 7 m/s is 55%, less then 8 m/s is 65% (fig.4).

## 3. Theory of Fabric Aerial Dams (FAD)

**1**. **Dynamic pressure** *P* of motion air (wind) can be computed by equation:

$$P = \frac{\rho V^2}{2}, \qquad\qquad (3)$$

here $P$ is wind dynamic pressure, N/m$^2$; $\rho = 1.225$ kg/m$^3$ (standard value) is air density;

$V$ is wind speed, m/s. Result of computation is presented in Fig. 5.

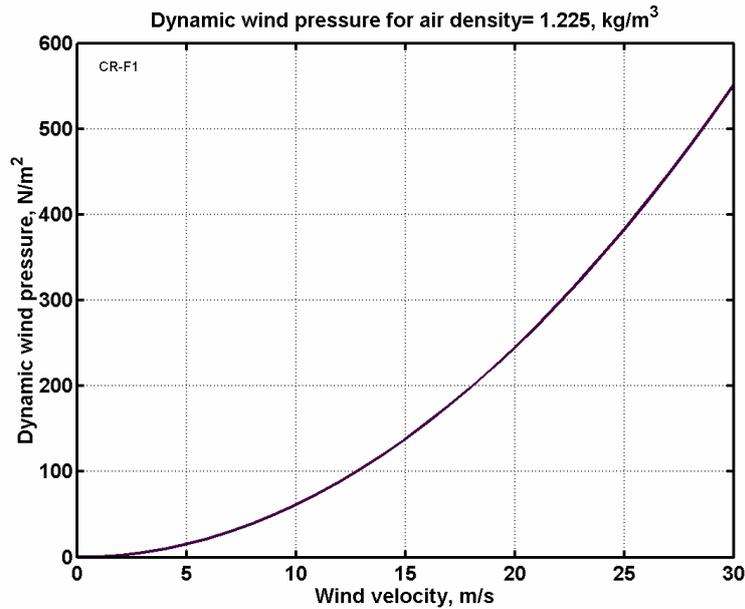

**Fig. 5.** Wind dynamic pressure via wind speed

**2**. **The wind power** can be computed by

$$N = \eta \frac{\rho A V^3}{2}, \qquad\qquad (4)$$

where $N$ is power, W; $\eta$ is coefficient efficiency of air turbine, $\eta = 0.3 \div 0.5$; $A$ is turbine area, m$^2$; a conical entry into turbine 4 is shown in Fig. 2 can increase the effective area sometimes.

The annual average wind speed near latitude 60$^0$ S is 12.7 m/s at height 50 m. If FAD has an off-the-ice cap height of 50 m and is 1,000 km long, a coefficient efficiency of 0.5, the total power of FAD wind turbines may be more than 30 GW. In other words, a ring-shaped FAD 'enclosing' Antarctica theoretically might generate 450 GW of electricity."

**3**. **Annual energy** *E* received from wind turbine is

$$E = 8.33N \quad \text{[kWh]} \ . \tag{5}$$

Computation of this equation is presented in fig. 6.

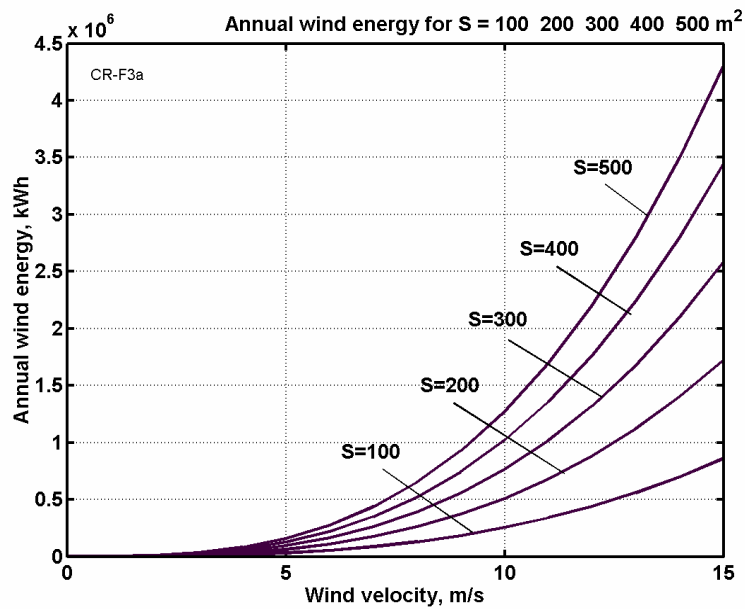

**Fig. 6**. Annual wind energy via wind velocity and turbine area *S*.

**4**. **Requisite film thickness** $\delta$ is

$$\delta = \frac{T}{2\sigma}, \quad T = PH \ , \tag{6}$$

where *T* is wind force in 1 m width of the film, N/m; *H* is dam height, m (fig. 2); $\sigma$ is safety tensile stress, N/m². Computational result for different values of $\sigma$ are presented in fig. 7-8. Multiplier 1/2 accounts the force *T* is kept in two points (top and bottom).

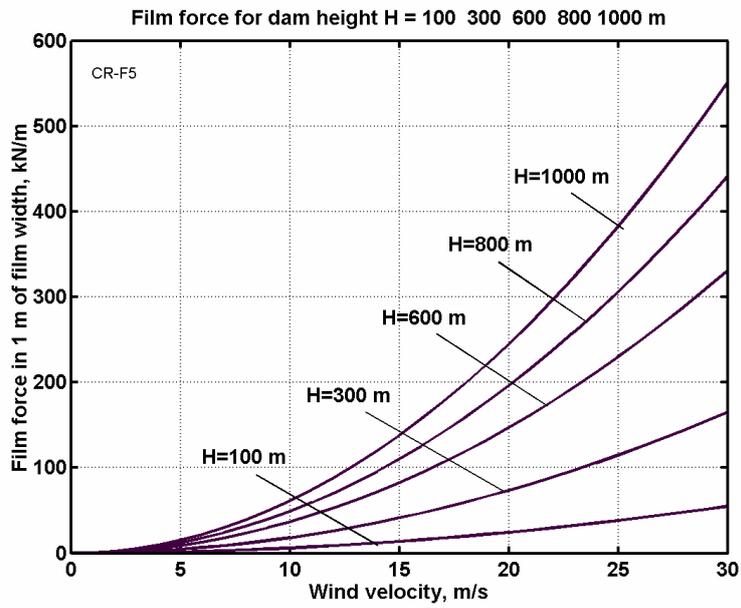

**Fig. 7**. Film force active in one meter film width versus wind velocity and height of dam *H*.

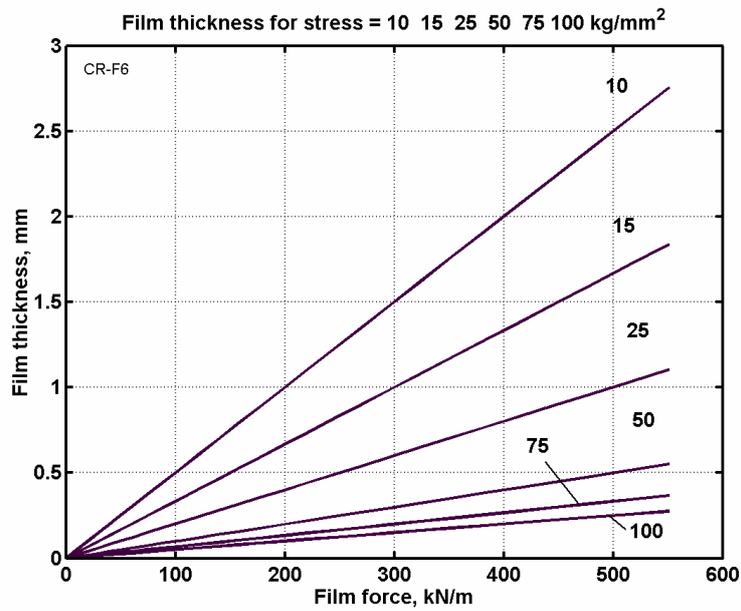

**Fig. 8**. Film thickness via film force and safety film stress

$$\sigma = 10, 15, 25, 50, 75, 100 \text{ kg/mm}^2.$$

## 5. The requisite diameter *d* of support cable is

$$S = \frac{T\,l_2}{\sigma}, \quad d = \sqrt{\frac{4S}{\pi}}, \tag{7}$$

where $S$ is cross-section area of cable [m$^2$]; $l_2$ is distance between cables, m. Results of computation for distance 10 m are presented in fig. 9. If distance between support cables is different from 10 m, the cross-section cable area must be changed in proportion.

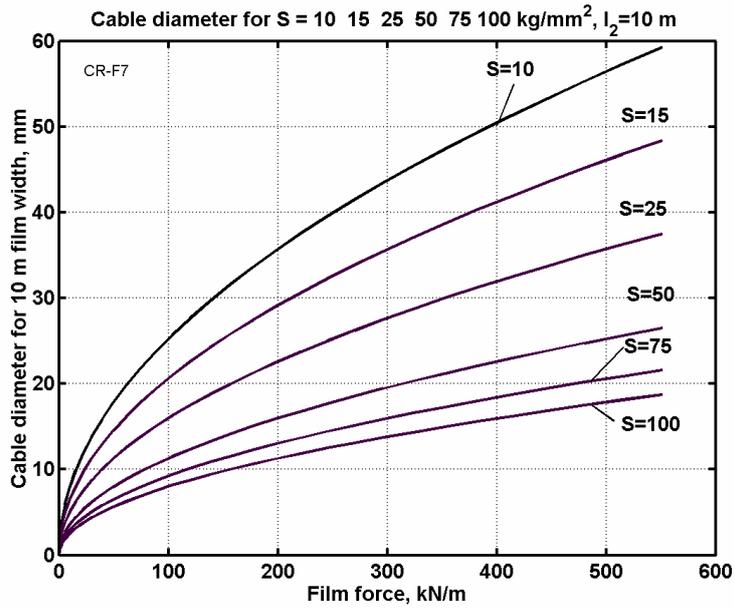

**Fig. 9**. Cable diameter for 10 m film width via film force and safety cable tensile stress.

**6. Weight $W_f$ of 1 m width film** is

$$W_f = 1.2H\delta\gamma, \tag{8}$$

here $\gamma$ is specific density of the film (conventionally, for the most artificial fiber $\gamma = 1800$ kg/m$^3$). Factor 1.2 take into curve form of film. Result of computation is in fig. 10.

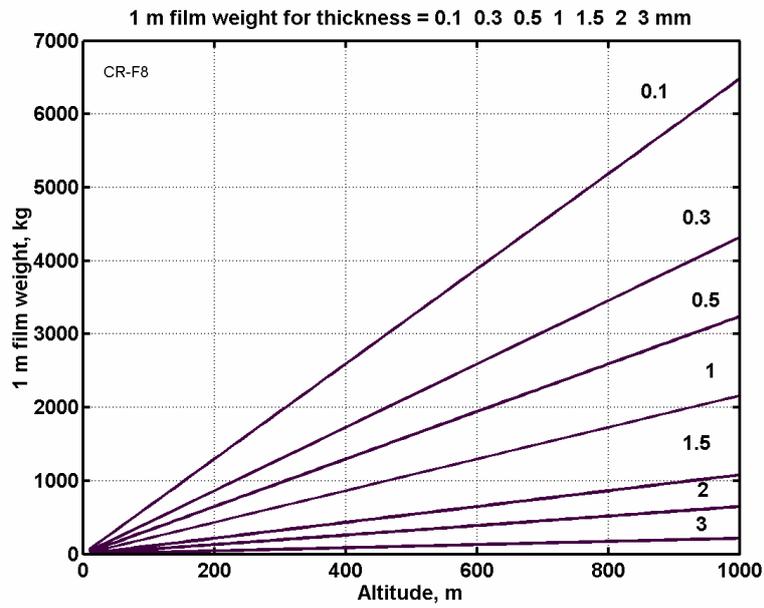

**Fig. 10**. Film weight of width 1 m via dam height and film thickness.

**7**. **The weight $W_c$ of the single cable** for angle 30$^\circ$ to horizon is

$$W_c = 2HS\gamma, \tag{9}$$

where $\gamma$ is specific density of the cable (conventionally, for the most artificial fiber $\gamma = 1800$ kg/m$^3$).

Results of computation are shown in fig. 11.

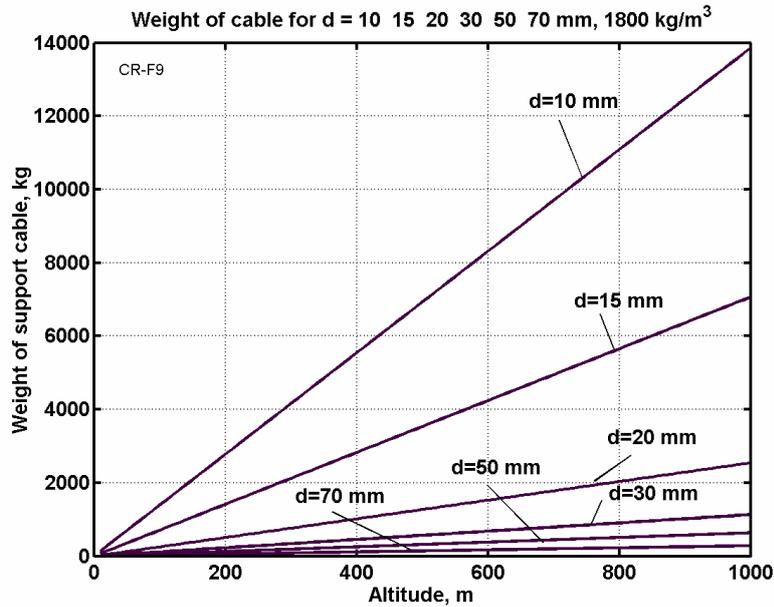

**Fig. 11**. Weight of film support cable versus dam height and cable diameter for cable density 1800 kg/m$^3$.

## 4. ENVIRONMENTAL CONSEQUENCES

All of the anchor points for the FAD tie-down cables will be in thick ice. We propose the use of "ice-anchors" that are implanted using a simple heated water drill far smaller than the device currently being used to reach a deep lake beneath Antarctica. These anchor points can be removed and reset without environmental consequence.

Isolated Antarctica, which surrounds the South Pole, has no native land-based vertebrates save flightless penguins. Since 2003, penguins have lived in close proximity with big wind turbines at the Australian Mawson Research Station; neither the tall metal tower structure nor the whirring of the large-diameter propellers seems to disturb their normal activities. Located on the coast, the Mawson Research Station is already harnessing the katabatic winds we intend to harness further inland. Other seabirds, such as the Wandering Albatross, the Grey Headed albatross, Antarctic Fulmers, Cormorants, Antarctic and Giant Petrels will not be affected by FAD mainly because these flying birds do not range inland very far. FAD will carefully be sited beyond their normal ocean feeding movements and nesting activities.

## 5. CONCLUSION

We have shown conclusively that Aerial Dams that include air turbines generating electricity can be successfully deployed and operated almost continuously on the continent of Antarctica. Construction of our macroproject would open vast territories to exploration and exploitation. The FAD, resembling the sails of watercraft, can be laced with conductive wires to heat slightly, which would amplify the Teflon-coated textile's capacity to shed snow and ice. Electric land vehicles could travel on safe paths, such as the Antarctic-1 highway, while drawing their motive power from battery-recharge stations fed by FAD. It seems reasonable to anticipate the use of battery-powered vehicle "Land Trains" such as an adaptation R.G. LeTourneau, Inc.'s 1955 SNO-FREIGHTER (Model VC-22), perhaps traveling periodically on an Antarctic-2 highway encircling Antarctica approximately 100 km inland from the coastline to maintain the FAD. Various industrial activities, such as mining remote ore bodies or petroleum deposits, will also be enhanced made more economic and convenient by the ready availability of large amounts of clean energy (Green electricity) derived from wind-power.

At present, the Protocol on Environmental Protection to the Antarctic Treaty, which came into force on 14 January 1998, precludes all extraction of minerals except for scientific study. However, with the increase of the world's population, new geopolitical pressures may soon arise that will initiate a major change of the Antarctic Treaty to accommodate the globalized needs of humanity.

Therefore, we anticipate an early-21st Century reopening of the worldwide debate on the Convention on Regulation of Antarctic Mineral Resource Activities that terminated in 1988.

We anticipate that Antarctica will, eventually, produce an excess of electricity and we, therefore, expect that such excess will be exported via super-conductive undersea electric power cables to South America.  Also, we forecast that FAD can be used to power the directed movement of Antarctica's  gigantic tabular icebergs to freshwater-short arid lands such as Australia (Husseinv, 1978). In additional to ground wind the Antarctic has strong high atmospheric flows. They also can be utilized (Bolonkin, 2004).

### *Cited References*